\documentclass{aa}  

\usepackage{natbib}
\usepackage{graphicx}
\usepackage{color} 

\titlerunning
\authorrunning

\bibpunct{(}{)}{;}{a}{}{,}
\usepackage{txfonts}
\begin{document}

   \title{ALMA Deep Field in SSA22:}

   \subtitle{A near-infrared-dark submillimeter galaxy at z=4.0}

   \author{
   Hideki Umehata\inst{1,2},
   Ian Smail\inst{3},
   A.M. Swinbank\inst{3},
   Kotaro Kohno\inst{2,4},
   Yoichi Tamura\inst{5},
   Tao Wang\inst{2,6},
   Yiping Ao\inst{7,8},
   Bunyo Hatsukade\inst{2},
   Mariko Kubo\inst{9,6},
   Kouchiro Nakanishi\inst{6,10},
   Natsuki N. Hayatsu\inst{6}
          }

   \institute{RIKEN Cluster for Pioneering Research, 2-1 Hirosawa, Wako, Saitama 351-0198, Japan\\
              \email{hideki.umehata@riken.jp}
         \and
             Institute of Astronomy, Graduate School of Science, The University of Tokyo, 2-21-1 Osawa,
Mitaka, Tokyo 181-0015, Japan
        \and
        Centre for Extragalactic Astronomy, Department of Physics, Durham University, South Road, Durham DH1 3LE, UK
        \and
Research Center for the Early Universe, Graduate
School of Science, The University of Tokyo, 7-3-1 Hongo, Bunkyo-ku,
Tokyo 113-0033, Japan
    \and
        Division of Particle and Astrophysical Science, Graduate School of Science, Nagoya University, Aichi 464-8602, Japan
        \and
        National Astronomical Observatory of Japan, 2-21-1, Osawa, Mitaka, Tokyo 181-8588, Japan
        \and
        Purple Mountain Observatory and Key Laboratory for Radio Astronomy, Chinese Academy of Sciences, Nanjing, China
        \and
        School of Astronomy and Space
Science, University of Science and Technology of China, Hefei, China
        \and
        Research Center for Space and Cosmic Evolution, Ehime University, Bunkyo-cho 2-5, Matsuyama 790-8577, Japan
        \and
        Department of Astronomical Science, SOKENDAI (The Graduate University for Advanced Studies), Mitaka, Tokyo 181-8588, Japan
             }
    \titlerunning{A near-infrared-dark galaxy at z=4}
    \authorrunning{H. Umehata et al.} 

   \date{Received MM DD, 2020; accepted MM DD, 2020}

 
  \abstract
{Deep surveys with the Atacama Large Millimeter Array (ALMA) have uncovered a population of dusty star-forming galaxies which are faint or even undetected at optical to near-infrared wavelengths. 
Their faintness at short wavelengths makes the detailed characterization of the population challenging. Here we present a spectroscopic redshift identification and a characterization of one of these near-infrared-dark galaxies discovered by an ALMA deep survey. The detection of [C~{\sc i}](1-0) and CO(4-3) emission lines determines the precise redshift of the galaxy, ADF22.A2, to be $z=3.9913\pm 0.0008$. On the basis of a multi-wavelength analysis, ADF22.A2 is found to be a massive, star-forming galaxy with a stellar mass of $M_*=1.1_{-0.6}^{+1.3}$~$\times10^{11} M_\odot$ and SFR$=430_{-150}^{+230}$~$M_\odot$~yr$^{-1}$. 
The molecular gas mass was derived to be $M (H_2)^{\rm [CI]}=(5.9\pm1.5)\times10^{10}$ $M_\odot$, indicating a gas fraction of $\approx35\%$, and the ratios of $L_{\rm [CI](1-0)}/L_{\rm IR}$ and $L_{\rm [CI](1-0)}/L_{\rm CO(4-3)}$ suggest that the nature of the interstellar medium in ADF22.A2 is in accordance with those of other bright submillimeter galaxies. The properties of ADF22.A2, including the redshift, star-formation rate, stellar mass, and depletion time scale ($\tau_{\rm dep}\approx0.1-0.2$~Gyr), also suggest that ADF22.A2 has the characteristics expected for the progenitors of quiescent galaxies at $z\gtrsim3$.
Our results demonstrate the power of ALMA contiguous mapping and line scan, which help us to obtain an unbiased view of galaxy formation in the early Universe.
}

   \keywords{submillimeter: galaxies --
   galaxies: starburst --
   galaxies: high-redshift --
   submillimeter: ISM --
   galaxies: evolution --
               }

   \maketitle
%

\section{Introduction}

\begin{figure}[h]
\centering    
\includegraphics[width=8.5cm]{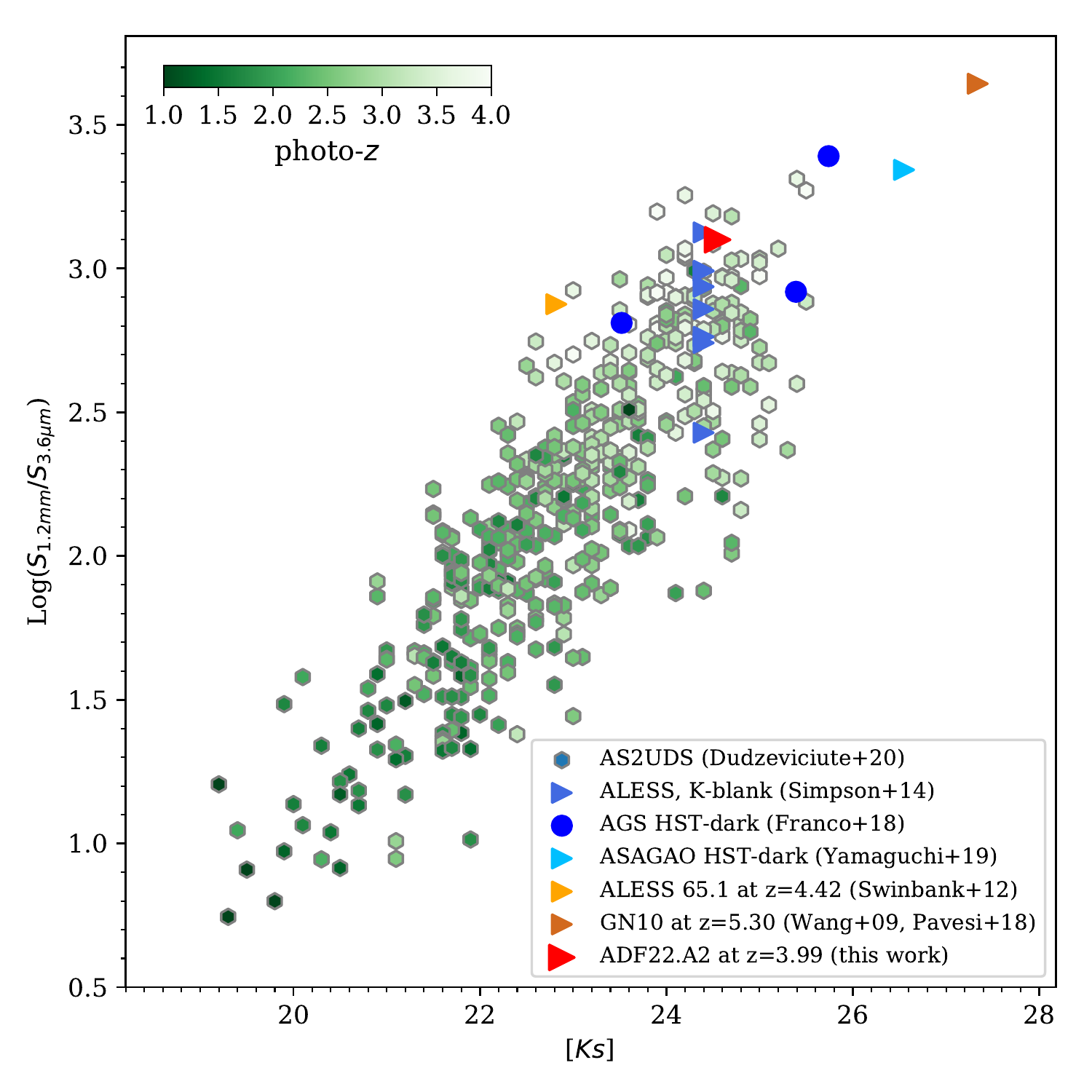}
    \caption{
Magnitude-color diagram of ADF22.A2 where the horizontal axis shows {\it Ks} band magnitude in the AB system, while the vertical axis shows the flux ratio $S_{\rm 1.2 mm}/S_{\rm 3.6\mu m}$ in a logarithmic scale. We also show {\it HST}-dark galaxies found by ALMA deep surveys (\citealt{2018A&A...620A.152F}; \citealt{2019ApJ...878...73Y}) and {\it K}-blank SMGs with $z_{\rm spec}$ (\citealt{2012MNRAS.427.1066S}; \citealt{2009ApJ...690..319W}; \citealt{2018ApJ...864...49P}). It is important to note that sources that do not have a reliable photometry due to source confusion were excluded. For comparison purposes, {\it K}-blank SMGs from \citet{2014ApJ...788..125S} and {\it K}-band detected SMGs in the
ALMA SCUBA-2 UDS survey (AS2UDS, \citealt{2020MNRAS.tmp.1074D}) are also displayed. We adopted 3$\sigma$ upper limits for {\it K}-blank sources.
ADF22.A2 shares a similar region with other near-infrared-dark ALMA sources, which demonstrates that they belong to an equivalent population. The $z_{\rm phot}$ distribution of SMGs in AS2UDS and some $z_{\rm spec}$ for the near-infrared-dark population suggests that the population is generally estimated to be at $z>3.5$ and ADF22.A2 is one example of them.}
\label{fig:sample}
\end{figure}

Intense star-forming activity in a growing galaxy is often obscured by dust in the early Universe. The absorption of stellar emission and reradiation by dust make the galaxy apparently faint at optical-to-near-infrared wavelengths even though they have star-formation rates of $\sim$ 100-1000~$M_\odot$~yr$^{-1}$ (\citealt{2002PhR...369..111B}; \citealt{2014PhR...541...45C} for reviews). Consequently it is challenging to recognize a fraction of star-forming galaxies selected at (sub)millimeter even in a deep near-infrared image. Such a "near-infrared-dark" population was first identified among bright submillimeter galaxies (SMGs) (e.g., \citealt{1999ApJ...519..610D}; \citealt{1999MNRAS.308.1061S}; \citealt{2004AJ....127..728F}). 
The faintness makes it difficult to determine spectroscopic redshift $z_{\rm spec}$ using traditional spectroscopic observations at (observed) optical-to-near infrared wavelengths, which has been a major obstacle in studying these populations. As an alternative, 
bright rest-frame far-infrared (FIR) lines such as [C~{\sc ii}] and CO lines has been utilized to determine $z_{\rm spec}$, which undoubtedly shows that they are heavily dust-obscured star-forming galaxies at high redshift. For instance, \citet{2009ApJ...690..319W} placed a strong upper limit ($K_s$=27.3~AB at 3$\sigma$) on GN10 at $z=5.303$ (the redshift has been reported by \citealt{2018ApJ...864...49P}), and \citet{2012Natur.486..233W} found that HDF850.1, which has no counterpart on an {\it HST} image, is located at $z=5.183$. 

Following these works in the era when the Atacama Large Millimeter/submillimeter Array (ALMA) had not been available,
the advent of interferometers such as ALMA has allowed us to detect and precisely locate fainter and more abundant (sub)millimeter populations. One avenue of research is follow-up observations of SMGs that were originally discovered by a single-dish telescope. For instance, \citet{2014ApJ...788..125S} found ten SMGs which were detected at 3.6~$\mu$m, but they were blank at shorter wavelengths. 
While obtaining a spectroscopic redshift of these types of near-infrared-dark SMGs is still not easy, 
a handful of these galaxies are now placed at $z\sim4-5$ via molecular  and atomic line detections at (sub)millimeter and the number is increasing
(e.g., \citealt{2012MNRAS.427.1066S}; \citealt{2017ApJ...850....1R}; \citealt{2018ApJ...856...72O}; \citealt{2019ApJ...887...55C}; \citealt{2019ApJ...887..144J};
\citealt{2020ApJ...890..171J};
\citealt{2020ApJ...895...81R}).

ALMA deep surveys, which use many pointings to create a contiguous field, are another avenue of research to measure dust-obscured star-forming activity across the cosmic time (e.g., \citealt{2015ApJ...811L...3T}; \citealt{2015ApJ...815L...8U}; \citealt{2016ApJ...833...67W}; \citealt{2017MNRAS.466..861D}). Some of the galaxies identified by these types of surveys are known to be blank at the {\it H}-band (``{\it HST}-dark''; e.g., \citealt{2018A&A...620A.152F}) and {\it Ks}-band (e.g., \citealt{2015ApJ...815L...8U}; \citealt{2016PASJ...68...82Y}; \citeyear{2019ApJ...878...73Y}).
The existence of the near-infrared-dark ALMA galaxies has also been uncovered using another approach. \citet{2019Natur.572..211W} report the identifications of ALMA sources which were originally selected as ``{\it H}-dropout'' galaxies, which are IRAC sources without {\it HST}/WFC3 F160W counterparts. They suggest that these types of ``ALMA-detected {\it H}-dropouts'' are the bulk populations of massive ($M_*\gtrsim10^{10.3}$ $M_\odot$) star-forming galaxies at $z\approx3-6$ and that they dominate the total star-formation rate density at such a massive regime in the era (see also \citealt{2020MNRAS.tmp.1074D}).

These works suggest that some fraction of galaxies could have been missed in previous studies on the basis of the optical-to-near-infrared selection. Charting the nature of the near-infrared-dark ALMA galaxies is necessary for a comprehensive view of galaxy formation and evolution in the early Universe.
While the faintness of the optical-to-near-infrared emission makes the spectroscopy in these types of bands quite challenging, the search for molecular and atomic lines at the (sub)millimeter wavelengths can be a solution.
In this letter, we report a secure redshift identification of a near-infrared-dark galaxy discovered by an ALMA deep survey. We derive properties of the interstellar medium (ISM) and discuss the role of these types of galaxies in the galaxy formation history. 
Throughout this letter, we adopt a cosmology with 
$\Omega_{\rm m}=0.3, \Omega_\Lambda=0.7$, and H$_0$=70 km s$^{-1}$ Mpc$^{-1}$.


\section{Observations}

\begin{figure}
\centering    
\includegraphics[width=8.5cm]{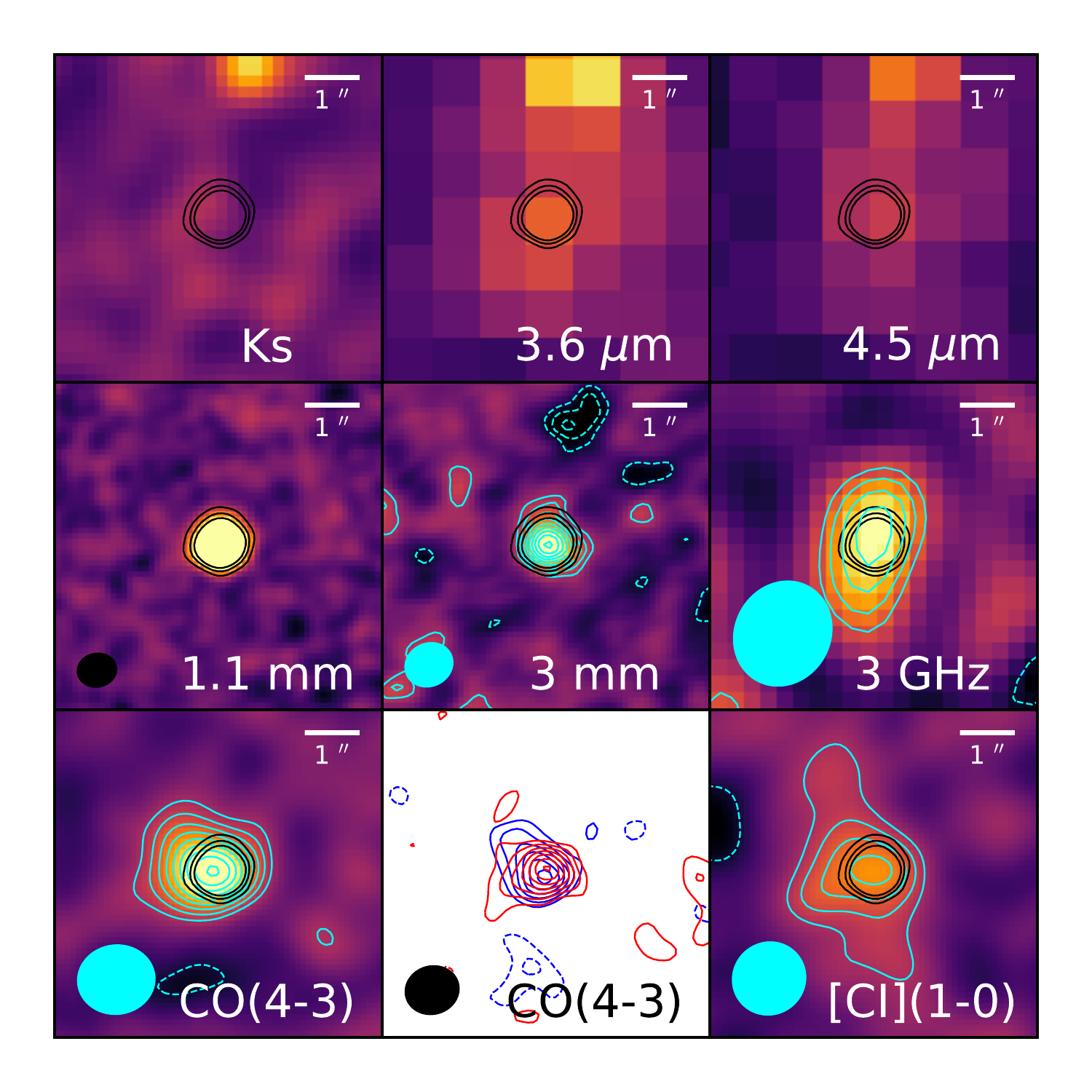}
    \caption{
Multiwavelength images of ADF22.A2. Black contours show 3$\sigma$, 6$\sigma$, and 9$\sigma$ of 1.1~mm emission. Cyan contours show 3 mm continuum, 3~GHz continuum, CO(4-3), or [C~{\sc i}](1-0) emission ($\pm2\sigma, \pm3\sigma$...) as denoted in each panel. ADF22.1 is blank at the {\it Ks}-band (and shorter wavelengths), which is a common feature of near-infrared-dark ALMA galaxies. We also show blue and red channels of CO(4-3) emission using contours ($\pm2\sigma, \pm3\sigma$...). The mean velocities of the two channels are $\pm200$~km~s$^{-1}$. The offset at the peak positions indicates a rotating gas disk.
}
\label{fig:stamp}
\end{figure}

Our target, ADF22.A2, was originally discovered through our ALMA deep survey in the SSA22 field. This field represents a significant proto-cluster at $z\approx3.1$ (e.g., \citealt{1998ApJ...492..428S}; \citealt{2004AJ....128.2073H}; \citealt{2012AJ....143...79Y}), and the ALMA mosaic was designed to observe a $6^\prime \times 3^\prime$ region located at the proto-cluster core at 1.1~mm (ALMA deep field in SSA22 or ADF22; \citealt{2015ApJ...815L...8U}; \citeyear{2017ApJ...835...98U}; \citeyear{2018PASJ...70...65U}; \citeyear{2019Sci...366...97U}; \citealt{2017PASJ...69...45H}). ADF22.A2 is the second brightest source in a $2^\prime \times 3^\prime$ subregion among ADF22 (ADF22A) with a flux density of $S_{\rm 1.1~mm}=2.02\pm0.02$~mJy (\citealt{2017ApJ...835...98U}). 
ADF22.A2 was detected at 3.6~$\mu$m and 4.5~$\mu$m, while it was completely invisible even at the {\it Ks}-band or shorter wavelengths (Table.~\ref{table:photometry}).
%
%
In this work, we refer to ALMA-identified galaxies, which have no detectable counterpart at optical-to-near-infrared wavelengths, as ``near-infrared-dark'' galaxies. ADF22.A2 is then an example of the population. While this practical selection allows us to shed light on the  populations which are missed in optical-to-near infrared surveys on the basis of an available data set, it is not straightforward to compare galaxies between different works since there are a wide variety of data sets, including bands and depths as well as the brightness of galaxies. Considering this ambiguity, 
 we compared ADF22.A2 with other near-infrared-dark ALMA galaxies (\citealt{2009ApJ...690..319W};
\citealt{2012MNRAS.427.1066S};
\citealt{2014ApJ...788..125S}; \citealt{2018A&A...620A.152F}; \citealt{2019ApJ...878...73Y}) and a systematical sample of ALMA-identified SMGs (\citealt{2020MNRAS.tmp.1074D}) in Fig. \ref{fig:sample}, where the {\it Ks}-band magnitude and a flux ratio (Log ($S_{1.2  \rm mm}/S_{3.6\mu \rm m}$)) are displayed.
ADF22.A2 and the majority of the near-infrared-dark populations are characterized as faint at the $Ks$-band ($\gtrsim24$ AB) and have a low $S_{\rm 1.2mm}$/$S_{\rm 3.6um}$ flux ratio ($\gtrsim 3$), which naively suggests that they are at a higher redshift and/or have a lower stellar mass among the ALMA-identified galaxies (see also \citealt{2014ApJ...788..125S}).

\begin{figure*}[h]
\centering    
\includegraphics[width=\hsize]{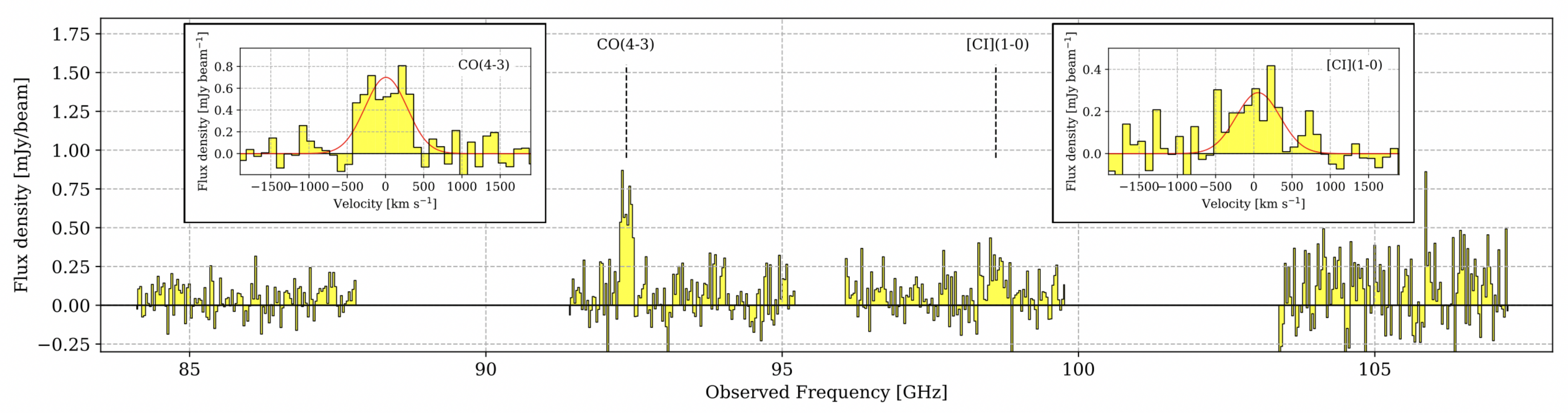}
    \caption{
    ALMA band~3 spectra of ADF22.A2 which covers CO(4-3) and [C~{\sc i}](1-0) lines. The inserted panels show spectra where velocities are relative to $z=3.991$.
Both lines show similar velocity profiles, which supports the idea that both lines trace molecular gas in the same galaxy - ADF22.A2.
    }
\label{fig:spectra}
\end{figure*}

Band~3 observations of ADF22.A2 were performed at two frequency set-ups. The first was carried out as a part of project 2015.1.00212.S and 2016.1.00543.S (PI. H. Umehata for both) at 84.1--87.8~GHz and 96.1--99.8~GHz (\citealt{2019Sci...366...97U}). The 36--44 available antennas were used and observations were executed in two periods; the former one was between 2016 July 21 and 2016 August 19, and the latter one was between 2017 May 7 and 2017 July 4. These resulted in the total on-source time of 13.9 hours for a 13-point mosaic, covering the 7 arcmin$^2$ area which is nearly cospatial with the 1.1~mm mosaic in ADF22A. 
J2226+0052 and J2148+0657 were observed for bandpass and phase calibration. To set the absolute flux scale, Pallas, J2148+0657, J2232+1143 and Titan were observed.
Subsequently, ADF22.A2 was observed as a part of project 2017.1.00602.S (PI. N. H. Hayatsu) at 91.4--95.2~GHz and 103.4--107.2~GHz. Observations were carried out with 43-45 usable 12 m antennas between 2018 January 18 and 20. The total on-source time was 263~min. 
The quasar J2148+0657 was utilized for the bandpass, phase, and flux calibration. 

The ALMA data were reduced using the Common Astronomy Software Application ({\sc casa}) versions 4.7.0, 4.7.2, 5.1.0, and 5.4.0 (\citealt{2007ASPC..376..127M}).
The data were processed using the standard {\sc casa} pipeline. 
The calibrated visibility data were used for imaging with the {\sc casa} task {\sc tclean}.
We created dirty cubes with a 100 km s$^{-1}$ velocity bin, applying {\sc uvtaper}, which yields the synthesized beam sizes of $\approx 1^{\prime\prime}.3$. The dirty cubes were utilized to identify emission lines. For the identified lines, cleaning was done down to $2\sigma$. As we show, we detected a bright CO(4-3) line, and hence we also repeated the clean process without any tapering for the spectral window which encompasses the line. 
A band~3 continuum map was also created using line-free channels without tapering, which have the synthesized beam $0^{\prime\prime}.90 \times 0^{\prime\prime}.71$ (P.A.$=-74$ deg) and a typical root mean square (RMS) noise of 44 $\mu$Jy beam$^{-1}$.


   \begin{table}
      \caption[]{Coordinate and multi-band photometry of ADF22.A2}
         \label{table:photometry}
         \begin{center}
            \begin{tabular}{c c }  
            \hline
            \noalign{\smallskip}
            RA (ALMA, ICRS) & 22~17~36.11     \\  
            Dec. (ALMA, ICRS) & 00~17~36.70    \\ 
            \noalign{\smallskip}
                        \hline
            $S_{1.14\rm{mm}}$ (ALMA) & 2.02 $\pm$ 0.02    \\
            $S_{3.13\rm{mm}}$ (ALMA) & $(59\pm9)\times10^{-3}$    \\
            \noalign{\smallskip}
                        \hline
            MegaCam 0.381 $\mu$m & $<8.3\times10^{-5}$    \\
            Suprime-Cam 0.437 $\mu$m & $<5.8\times10^{-5}$    \\
            Suprime-Cam 0.545 $\mu$m & $<5.2\times10^{-5}$    \\
            Suprime-Cam 0.651 $\mu$m & $<4.8\times10^{-5}$   \\
            Suprime-Cam 0.768 $\mu$m & $<6.3\times10^{-5}$   \\
            Suprime-Cam 0.920 $\mu$m & $<1.2\times10^{-4}$ \\
            MOIRCS 1.25 $\mu$m & $<4.2\times10^{-4}$    \\
            MOIRCS 1.63 $\mu$m & $<6.6\times10^{-4}$    \\
            MOIRCS 2.15 $\mu$m & $<5.5\times10^{-4}$    \\
            IRAC 3.6 $\mu$m & $(1.0\pm0.2)\times10^{-3}$    \\
            IRAC 4.5 $\mu$m & $(1.3\pm0.2)\times10^{-3}$    \\
            IRAC 5.6 $\mu$m & $<4.7\times10^{-3}$    \\
            IRAC 8.0 $\mu$m & $<6.3\times10^{-3}$    \\
            SPIRE 250 $\mu$m & $<11\times10^{-3}$    \\
            SPIRE 350 $\mu$m & $<12\times10^{-3}$    \\
            SPIRE 500 $\mu$m & $<16\times10^{-3}$    \\
            SCUBA2 850 $\mu$m & 3.67 $\pm$ 1.18    \\
            JVLA 3 GHz  & $(11\pm2)\times10^{-3}$   \\
            \noalign{\smallskip}
            \hline
         \end{tabular}
         \end{center}
         \tablefoot{Fluxes in units of mJy that are shown and all limits are 3$\sigma$. MegaCam $U$-band, Suprime-Cam $B, V, R, i^{\prime},z^{\prime}$-band, MOIRCS $J, H, Ks$-band, and IRAC data are summarized in \citet{2014MNRAS.440.3462U}. We also use data taken by SPIRE  (\citealt{2016MNRAS.460.3861K}), SCUBA2 (\citealt{2017MNRAS.465.1789G}), and JVLA (\citealt{2017ApJ...850..178A}).
         }
   \end{table}

\section{Results}

\subsection{Line detections and spectroscopic redshift}

As is shown in Fig.~\ref{fig:stamp} and Fig.~\ref{fig:spectra}, we successfully detected two emission lines in the combined band~3 spectra at around 92.4~GHz and 98.6~GHz, which provides an unambiguous redshift of $z_{\rm spec}=3.9913\pm0.0008$ for CO(4-3) ($\nu_{\rm rest}=461.041$~GHz) and [C~{\sc i}](1-0) ($\nu_{\rm rest}=492.160$~GHz). The redshift was determined from a one-component Gaussian profile fitting to the CO(4-3) spectra, which has a superior signal-to-noise ratio (S/N), while the [C~{\sc i}](1-0) line profile shows a consistent value $z_{\rm [CI]}=3.9921\pm0.0012$. Line fluxes were primary estimated from the Gaussian profile fit to the spectra extracted at a centroid position in an integrated line-emission map. We also measured the CO(4-3) flux on the integrated emission map using the {\sc casa} task {\sc imfit} to find a slightly (16\%) larger value. We finally adopted fluxes of $I_{\rm CO(4-3)}=0.56\pm 0.09$ Jy km s$^{-1}$ and $I_{\rm [CI](1-0)}=0.24\pm 0.06$ Jy km s$^{-1}$, while applying the aperture correction. These estimates correspond to line luminosities $L_{\rm CO(4-3)}=(0.69\pm 0.11$) $\times10^8$ $L_\odot$ and $L_{\rm [CI](1-0)}=(0.31\pm 0.08$) $\times10^8$ $L_\odot$, respectively.
We note that the brightness distributions of the [C~{\sc i}](1-0) and CO(4-3) emission lines can decrease due to the effect of the cosmic microwave background (CMB) at $z\approx4$ (\citealt{2013ApJ...766...13D}; \citealt{2016RSOS....360025Z}), though it is difficult to properly estimate the effect on ADF22.A2.

\subsection{Stellar mass, IR luminosity, and SFR}

\begin{figure}
\centering    
\includegraphics[width=8.5cm]{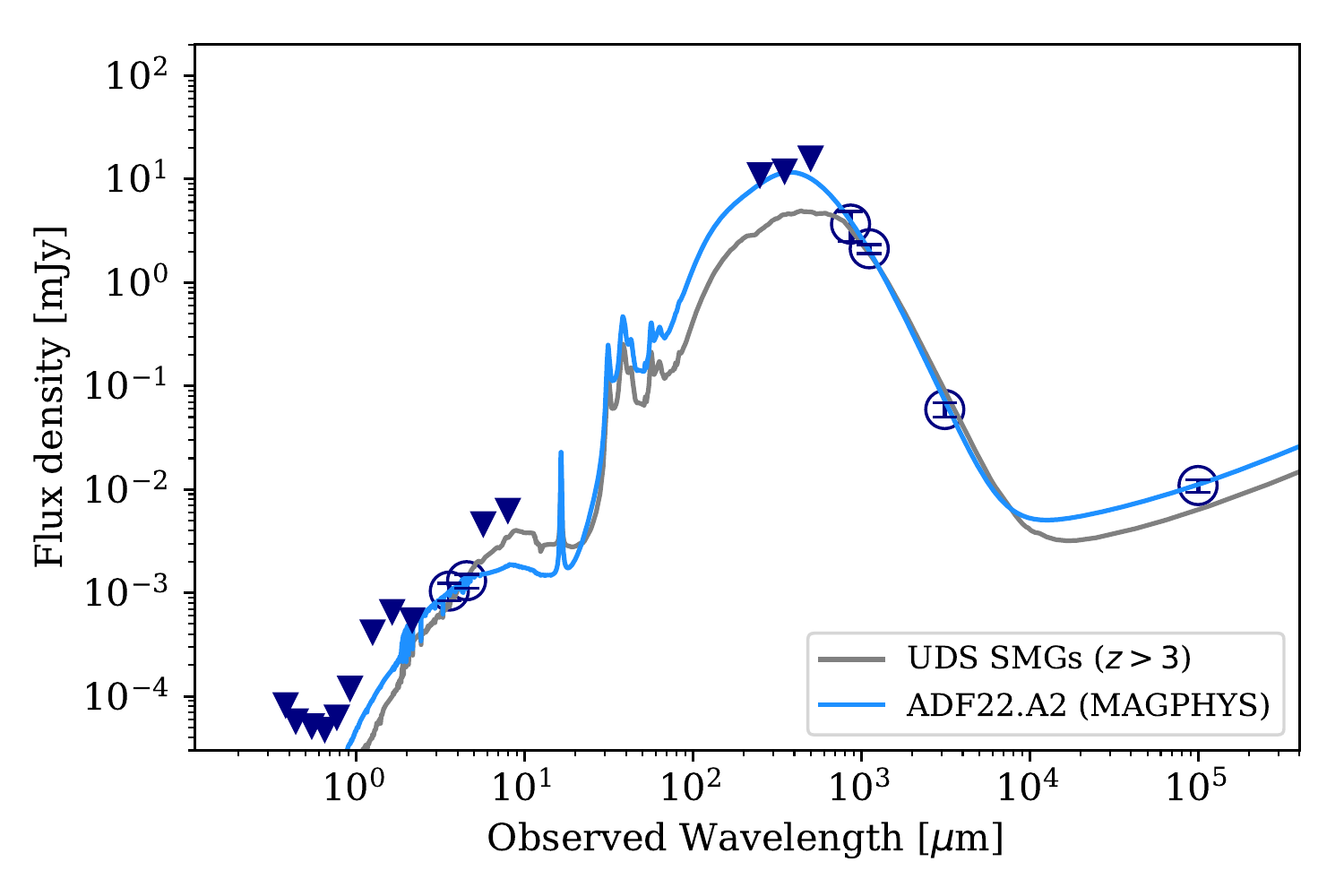}
    \caption{
Photometric points of ADF22.A2 and the best-fit SED by {\sc magphys}. Filled triangles show the upper limits. For comparison purposes, we also show averaged SEDs for UDS SMGs at $z_{\rm phot}>3$, which are scaled at 1.1~mm and assumed to be at $z=3.991$ (\citealt{2020MNRAS.tmp.1074D}).  The faint nature at the {\it K}-band and shorter wavelengths can also likely be explained with the $z>3$ template, while the limited number of available photometries causes some uncertainties for derived parameters.
    }
\label{fig:sed}
\end{figure}

To infer several physical parameters making use of the precise redshift, we applied the high-redshift extension of the SED-fitting algorithm {\sc magphys}
which includes the effects of dust attenuation (\citealt{2000ApJ...539..718C}) and describes star formation history by a range of ages and star formation time scales (\citealt{2008MNRAS.388.1595D}; \citeyear{2013ApJ...765....9D}).
We used the available photometry, which cover optical-to-near infrared (from the {\it U} band to {\it Spitzer} IRAC 8~$\mu$m; \citealt{2014MNRAS.440.3462U} and references therein), {\it Herschel}/SPIRE bands (\citealt{2016MNRAS.460.3861K}), SCUBA2 (\citealt{2017MNRAS.465.1789G}), ALMA band~6 (\citealt{2017ApJ...835...98U}), band~3, and JVLA 3~GHz (\citealt{2017ApJ...850..178A}). The 3~mm continuum flux was measured to be $S_{\rm 3.13mm}=59.2\pm9.3$~$\mu$Jy using {\sc casa}/{\sc imfit}. While we measured flux densities at the optical to near-infrared through aperture photometries in the same manner as is reported in  \citet{2014MNRAS.440.3462U}, the counterpart in IRAC 3.6~$\mu$m and 4.5~$\mu$m images is close to a foreground object and thus we estimated the fluxes deblending the two using GALFIT (\citealt{2002AJ....124..266P}) (see Appendix for details). The SPIRE fluxes were also deblended in same manner as in \citet{2014MNRAS.438.1267S}. 
The photometry and best-fit SED are shown in Fig.~\ref{fig:sed} and Table.~\ref{table:photometry}. 
For comparison purposes, averaged SEDs, which were derived for $z>3$ SMGs in the UDS field using the same code, {\sc magphys}, and scaled at 1.1~$\mu$m flux, are displayed (\citealt{2020MNRAS.tmp.1074D}). A large fraction of the ADF22.A2 photometry is also explainable by the template, which indicates that ADF22.A2 may have similar characteristics as the other SMGs at $z>3$.

We note that ADF22.A2 has no X-ray counterpart (\citealt{2009MNRAS.400..299L}) and thus the existence of an X-ray active galactic nucleus (AGN) has not been identified. The 3~GHz radio continuum emission is also well fitted by {\sc magphys}, which suggests ADF22.A2 show a general, far-infrared correlation, and are not associated with a radio-loud AGN.
Due to the relatively poor constraints around the peak of the SED, the derived dust properties might be more uncertain. 
The effect of the CMB on the continuum flux in the Rayleigh-Jeans regime (e.g., \citealt{2013ApJ...766...13D}; \citealt{2019ApJ...887..144J}) is also not considered in this paper.

The fitting yields median values and 68\% confidence intervals for the stellar mass $M_*=1.1_{-0.6}^{+1.3}$~$\times10^{11} M_\odot$, the infrared luminosity $L_{\rm IR:8-1000 \mu m}=5.5_{-1.7}^{+3.0}$~$\times10^{12}$~$L_\odot$, and the star-formation rate $430_{-150}^{+230}$~$M_\odot$~yr$^{-1}$.
We note that the relatively large uncertainties principally arose due to the limited number of available photometry points. The faintness at optical-to-near-infrared wavelengths, including the non-detection at the restframe around 1.6~$\mu$m, hampered us from putting a constraint on the $A_{\rm V}$ and age as well as the stellar mass. The peak (or shorter wavelength) regime of the dust SED is only constrained by limits and hence it is difficult to constrain the related properties. The initial mass function (IMF) can also have an impact these derived properties. We adopted a Chabrier IMF (\citealt{2003PASP..115..763C}) (but see also \citealt{2018Natur.558..260Z}).


%
%

\subsection{Molecular gas mass and gas mass conversion factor}

The total molecular gas mass stored in ADF22.A2 was also estimated in three ways.
First, we utilized the [C~{\sc i}](1--0) emission as a tracer of the gas mass. The [C~{\sc i}](1--0) line is proposed to be a measure of the molecular gas mass. The critical density is similar to that of CO(1--0).
Moreover [C~{\sc i}] is sensitive to CO-dark molecular gas where cosmic rays can destroy CO (e.g., \citealt{2004ApJ...615L..29P}; \citealt{2017MNRAS.466.2825B}; \citealt{2017ApJ...840L..18J}). 
We calculated the [C~{\sc i}]-based H$_2$ mass, assuming an optically thin case, following \citet{2004ApJ...615L..29P}:

\begin{equation}
\begin{split}
M({\rm H_2})^{\rm [CI]}&=1375.8 D_{\rm L}^2(1+z)^{-1}\left(\frac{X_{\rm [CI]}}{10^{-5}}\right)^{-1}\left(\frac{A_{10}}{10^{-7} {\rm s}^{-1}}\right)^{-1} \\
 &\quad\times Q_{10}^{-1}S_{\rm [CI]}\Delta v
\end{split}
,\end{equation}
where $X_{\rm [CI]}$ is the abundance ratio of [C~{\sc i}]/H$_2$. Here we adopt $X_{\rm [CI]}=3\times10^{-5}$ (\citealt{2004ApJ...615L..29P}; \citealt{2017MNRAS.466.2825B}). The Einstein $A$ coefficient $A_{10}=7.93\times10^{-8}$ s$^{-1}$ was incorporated. 
We used the excitation factor $Q_{10}=0.6$ (\citealt{2017MNRAS.466.2825B}; \citealt{2018ApJ...856...72O}).
Then Equation (1) gives $M (H_2)^{\rm [CI]}=(5.9\pm1.5)\times10^{10}$ $M_\odot$.
Second, we also estimated the total H$_2$ mass using CO(4--3), though caution should be exercised since such a mid-$J$ transition line traces relatively dense and warm molecular gas. If we assume a brightness temperature ratio of $r_{43/10}=0.41$ and the conversion factor $\alpha_{\rm CO}$=1.0 $M_\odot$ (K km s$^{-1}$pc$^{2}$)$^{-1}$ (\citealt{2011MNRAS.412.1913I}; \citealt{2011ApJ...739L..31R}; \citealt{2013MNRAS.429.3047B}), the observed CO(4-3) line intensity suggests a molecular gas mass of $M (H_2)^{\rm CO(4-3)}=(5.4\pm1.2)\times10^{10}$~$M_\odot$.
Thirdly, we made use of the dust mass as a tracer. The dust mass derived by SED fitting using {\sc magphys} is $M_{\rm d}=4.3\pm0.9\times10^8$ $M_\odot$. Assuming a dust-to-gas mass conversion factor of 120 (\citealt{2008ApJS..178..189W}), this implies a molecular gas mass of $M (H_2)^{\rm dust}=(5.2\pm1.1)\times10^{10}$~$M_\odot$.
Thus estimates using the three methods are consistent. In this paper, we use the [C~{\sc i}]-based gas mass in the following discussion. 
Finally we checked a diversity of derived  $\alpha_{\rm CO}$, which brought the derived molecular gas mass with different tracers into agreement. The measurements on the basis of [C~{\sc i}] and dust give $\alpha_{\rm CO}=1.1\pm0.4$ and $\alpha_{\rm CO}=1.0\pm0.3$ through the comparison with CO-based gas mass, respectively. 
Thus the conversion factors derived from three clues are generally in good agreement for ADF22.A2, while there are a number of assumed parameters including $X_{\rm [CI]}$ and a dust-to-gas mass conversion.

\section{Discussion and conclusion}

\begin{figure*}[h]
\centering    
\includegraphics[width=17cm]{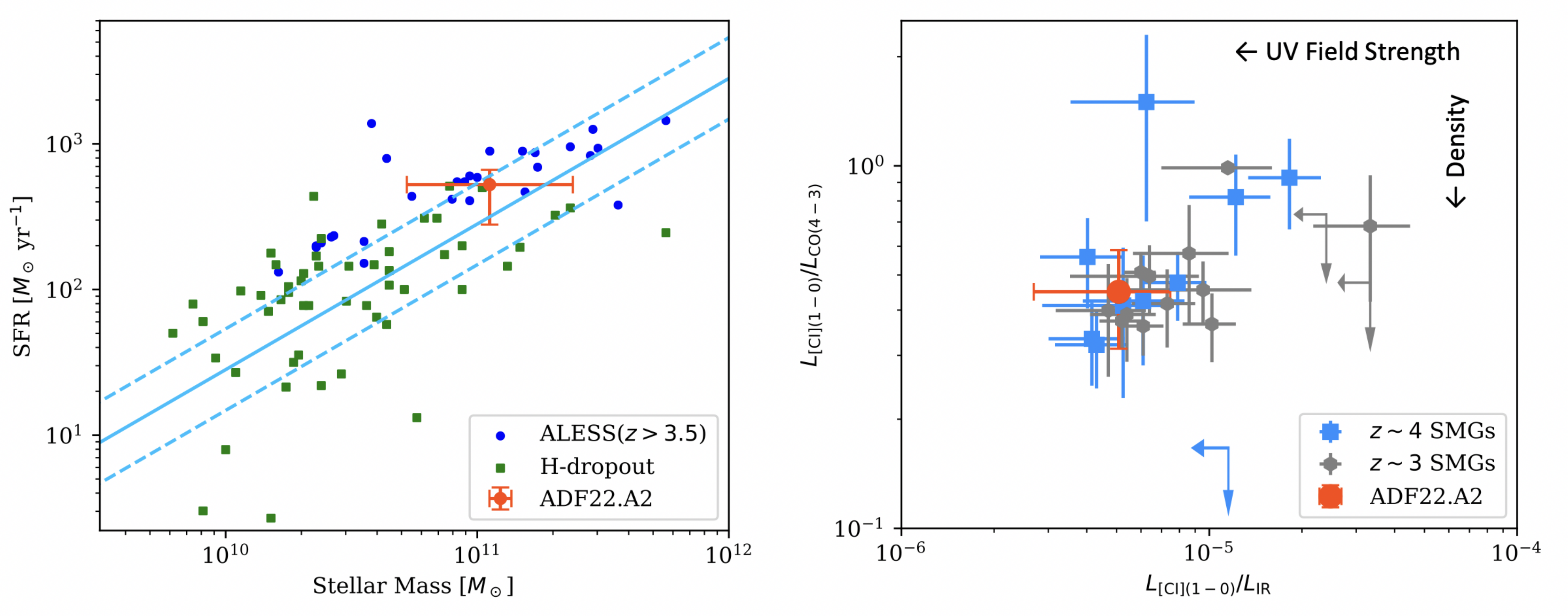}
    \caption{
(Left)
Relation between stellar mass and SFR for ADF22.A2, ALESS SMGs at $z>3.5$ (\citealt{2015ApJ...806..110D}), and {\it H}-dropout galaxies (\citealt{2019Natur.572..211W}). Cyan lines show the main-sequence of $z\sim4$ galaxies (\citealt{2017A&A...599A.134S}). ADF22.A2 is located at a region that is overlapped by the ALESS SMGs and H-dropout galaxies. 
(right)
[C~{\sc i}](1-0)/CO(4-3) Line luminosity ratio as a function of [C~{\sc i}](1-0) line and IR luminosity ratio for ADF22.A2 and SMGs at $z\sim3$ (\citealt{2013MNRAS.435.1493A};\citealt{2018A&A...620A..61C};\citealt{2019A&A...624A..23N}) and SMGs at $z\sim4$ (\citealt{2017MNRAS.466.2825B}). ADF22.A2 is located at the same region where the SMGs distribute in the diagram.
    }
\label{fig:nature}
\end{figure*}

The robust identification of its redshift, $z=3.991$, and the suite of multi-wavelength data sets pave the way to identify an example of near-infrared-dark galaxies in the history of galaxy assembly and understand the physical and chemical properties of these types of populations.
%
The left panel of Fig.~\ref{fig:nature} shows the relation between the stellar mass and SFR for ADF22.A2, compared with two other samples of near-infrared-dark galaxies and the  ``main-sequence'' relation at $z=4$ derived by \citet{2017A&A...599A.134S} on the basis of ALMA band~7 observations. 
The panel suggests that ADF22.A2 is characterized by massive dusty star-forming galaxies, which are located around (or above) the main-sequence at $z\sim4$.
ADF22.A2 is also located at a region where $z>3.5$ SMGs selected by an ALMA Survey of Submillimetre Galaxies in the Extended Chandra Deep Field South (ALESS, \citealt{2015ApJ...806..110D}) and "{\it H}-dropout" populations (\citealt{2019Natur.572..211W}), which tend to be fainter at (sub)millimeter than the SMGs originally sampled as a single-dish source overlap, which indicates that ADF22.A2 shares the properties with both populations. This would be reasonable, considering the fact that ADF22.A2 was first identified by an ALMA deep survey (\citealt{2015ApJ...815L...8U}) and successively identified by SCUBA2 (\citealt{2017MNRAS.465.1789G}), while it was missed by the previous AzTEC/ASTE survey (\citealt{2014MNRAS.440.3462U}).

Since ADF22.A2 was found in an ALMA deep survey which covers a significant cosmic volume contiguously, one can estimate the cosmic volume of the near-infrared-dark ALMA galaxies.
\citet{2019Natur.572..211W} estimated the number density of the {\it H}-dropout galaxies $\approx2\times10^{-5}$ Mpc$^{-3}$ on the basis of photometric redshifts. If we consider a similar redshift range ($z\approx3.4-5.4$) as was adopted in \citet{2019Natur.572..211W}, the cosmic volume covered by the $2^\prime \times 3^\prime$ region of ADF22A is $\approx3.5\times10^4$ Mpc$^{3}$, and the expected number of these types of populations is around unity. Considering the fact that the detection limit is comparable between the works (\citealt{2017ApJ...835...98U}; \citealt{2019Natur.572..211W}) and that all of the 18 ADF22A sources now have spectroscopic redshifts
(\citealt{2019Sci...366...97U}), both estimates are consistent.
This gives another clue supporting the prevalence of the near-infrared-dark ALMA galaxies in the early Universe. We also note that there are some caveats, however. The sample in  \citet{2019Natur.572..211W} is for {\it H}-dropout galaxies, and the selection criteria are not perfectly equal to ours. Most of the spectroscopic redshifts of the {\it H}-dropout galaxies are still lacking; hence, they leave  uncertainties as to the cosmic volume. 



\begin{figure}
\centering    
\includegraphics[width=7.5cm]{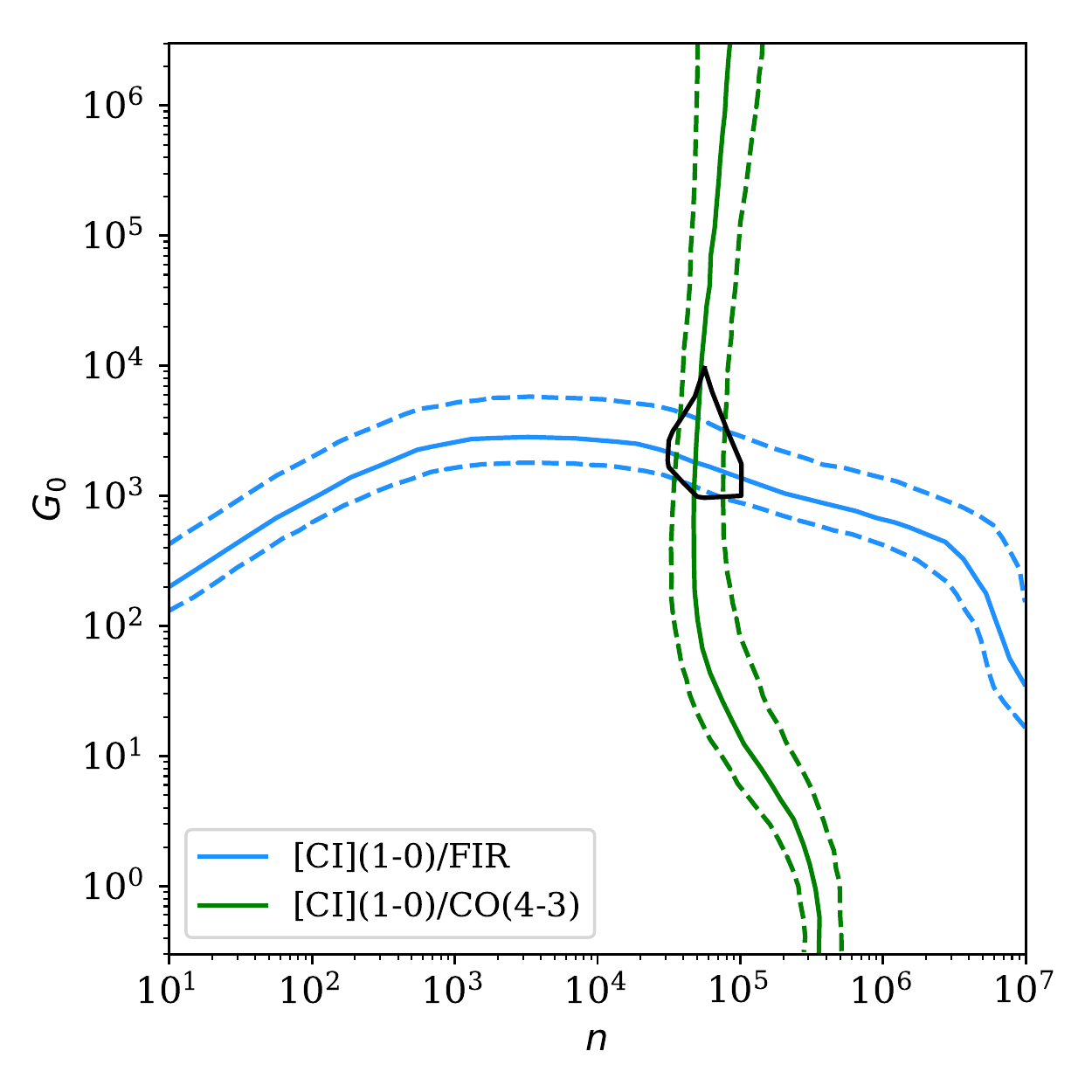}

    \caption{
Green and blue contours show $L_{\rm [CI](1-0)}/L_{\rm CO(4-3)}$ and $L_{\rm [CI](1-0)}/L_{\rm FIR}$ of ADF22.A2 on the gas density ($n$) and radiation field ($G_0$) plane from the PDR model of \citet{1999ApJ...527..795K} and \citet{2006ApJ...644..283K}. The black contour corresponds to the 68\% confidence levels.
}
\label{fig:pdr}
\end{figure}

The detection of the [C~{\sc i}](1--0) and CO(4--3) lines, together with infrared luminosity, gives us a clue as to the nature of the ISM.
We derived the gas density and the strength of the far-ultraviolet radiation field on the basis of the PDR model of \citet{1999ApJ...527..795K} and \citet{2006ApJ...644..283K}. While dust continuum and [C~{\sc i}](1--0) emission are generally optically thin, CO emission is optically thick. We increase the observed CO luminosity by a factor of two to incorporate this effect. Fig.~\ref{fig:pdr} shows the comparison between the observed ratios and the theoretical models.
The PDR models suggest the atomic gas density of $n=(5.6_{-2.5}^{+4.6})\times10^4$~cm$^{-3}$ and the UV radiation field $(1.8_{-0.8}^{+7.8})\times10^3~G_0$ for ADF22.A2 (Fig. \ref{fig:pdr}). 
The right panel of Fig. \ref{fig:nature} shows the ratios of $L_{\rm [CI](1-0)}/L_{\rm IR}$ and $L_{\rm [CI](1-0)}/L_{\rm CO(4-3)}$ compared with $z\sim4$ lensed SMGs (\citealt{2017MNRAS.466.2825B}; \citealt{2018A&A...620A..61C};\citealt{2019A&A...624A..23N}) and $z\sim3$ SMGs (\citealt{2013MNRAS.435.1493A}).
The consistency of the line ratios suggests that these characteristics, the UV radiation field strength and gas density of the ISMs within ADF22.A2, are similar to those of other dusty starburst galaxies selected from traditional bright SMG surveys.
More samples of near-infrared-dark ALMA galaxies, which have measurements of [C~{\sc i}](1--0) and CO(4--3) lines, will pave the way to further characterize ISM properties of the population.


We also placed ADF22.A2 in the context of galaxy evolution across the cosmic time.
Recently an increasing number of quiescent galaxies has been identified at $z\sim3-4$ (e.g., \citealt{2018A&A...618A..85S}; \citealt{2019ApJ...885L..34T}; \citealt{2019arXiv190910540V}). These findings invoke the prevalent galaxy populations which experience an active star-forming phase at higher redshifts. 
The measurements of ADF22.A2 suggest a gas-mass fraction of $\sim35\%$, which are defined as $f_{\rm gas}=M_{\rm gas}$/($M_{\rm gas}+M_*$). Together with a significant amount of stellar mass $M_*\approx10^{11}$~$M_\odot$, the gas mass fraction indicates that ADF22.A2 is a relatively evolved system.
The gas depletion time of ADF22.A2 is $\tau_{\rm dep}=M_{\rm gas}/SFR\approx0.1-0.2$~Gyr; hence, ADF22.A2 would evolve into a massive (log($M_*/M_\odot \approx 10^{11}$)) quiescent system at $z=3.6-3.8$ on the simple assumption that the galaxy simply consumes the estimated molecular gas reservoir with the constant star-formation rate. 
These results support the scenario that the near-infrared-dark ALMA populations are progenitors of quiescent galaxy populations at $z\approx3-4$.

ADF22.A2 demonstrates that ALMA deep surveys are a powerful tool to obtain an unbiased picture of galaxy formation and evolution, exemplifying an intriguing galaxy which was identified and characterized by ALMA observations. Further
 systematic census will help us understand the general nature of the near-infrared-dark ALMA galaxies.


\begin{acknowledgements}
We thank an anonymous referee for a number of constructive comments.
We thank Ms. Ugne Dudzevi\v{c}i\={u}t\.{e} for providing catalogs and a SED template.
We thank Prof.~R.~J.~Ivison for useful discussions.
H.U. and K.K. acknowledge support from JSPS KAKENHI grant (17K14252, 20H01953, 17H06130).
This work was supported by the NAOJ ALMA Scientific Research Grant Number 2017-06B.
I.R.S. and A.M.S. acknowledge support by the Science and Technology Facilities Council [grant number ST/P000541/1].
Y.A. acknowledges financial support by NSFC grant 11933011.
This paper makes use of the following ALMA data: ADS/JAO. ALMA\#2013.1.00162.S, \#2015.1.00212.S, \#2016.1.00543.S, \#2017.1.00602.S. ALMA is a partnership of ESO (representing its member states), NSF (USA) and NINS (Japan), together with NRC (Canada), MOST and ASIAA (Taiwan), and KASI (Republic of Korea), in cooperation with the Republic of Chile. The Joint ALMA Observatory is operated by ESO, AUI/NRAO and NAOJ.
The National Radio Astronomy Observatory is a facility of the National Science Foundation operated under cooperative agreement by Associated Universities, Inc.
This paper is based on data collected at
Subaru Telescope, which is operated by the National Astronomical Observatory of Japan. This work is based on observations made with the Spitzer Space Telescope, which is operated by the Jet Propulsion Laboratory, California Institute of Technology under a contract with NASA.
\end{acknowledgements}


\bibliographystyle{aa}
\bibliography{38146_arxiv1.bbl}

\begin{appendix}

\section{Deblending IRAC photometry}
\begin{figure}
\centering    
\includegraphics[width=7.5cm]{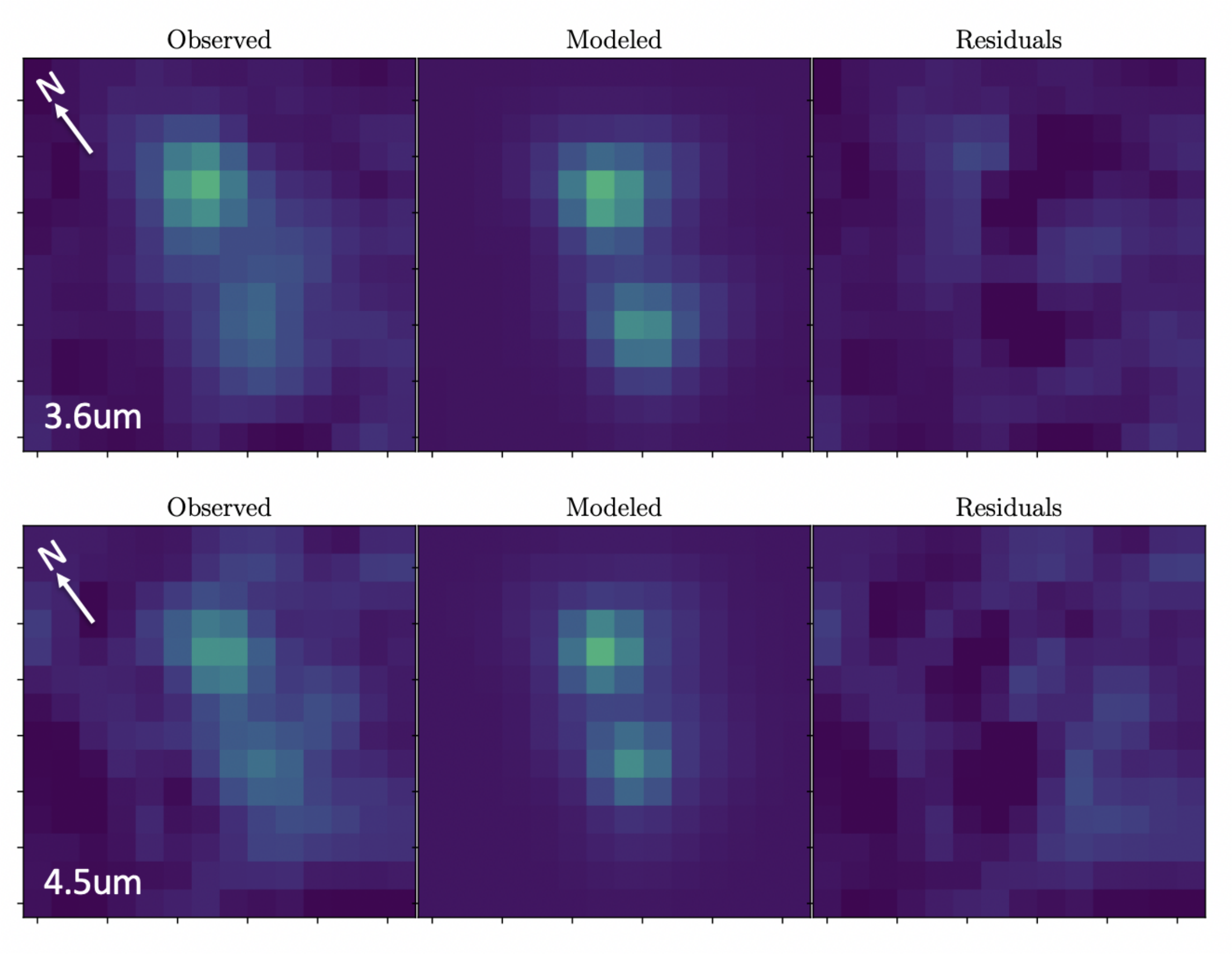}
    \caption{
 3.6~$\mu$m and 4.5~$\mu$m images of ADF22.A2. The left panels show observed images (the images are not rated, and this allows us to show the direction of the north.). The best-fit models are shown in the middle panels, and the residuals are shown in the right panel. Each is $9^{\prime\prime}\times9^{\prime\prime}$ in size.}
\label{fig:galfit}
\end{figure}

Since the counterpart of ADF22.A2 has a close companion in the 3.6~$\mu$m and 4.5~$\mu$ images, we measured the flux density using {\sc galfit} (\citealt{2002AJ....124..266P}). We fit the region around ADF22.A2 using three components, including two point sources and the sky. Fig.~\ref{fig:galfit} shows the best-fit results. The derived fluxes are $S_{\rm 3.6\mu m}=1.0\pm0.2$~$\mu$Jy and $S_{\rm 4.5\mu m}=1.3\pm0.2$~$\mu$Jy, respectively (see also Table~\ref{table:photometry}). These measurements exceed the 3$\sigma$ detection limits for a point source (0.4~$\mu$Jy and 0.8~$\mu$Jy at 3.6~$\mu$m and 4.5~$\mu$m, respectively), as was derived by \citet{2009ApJ...699.1610H}.

\end{appendix}

\end{document}